\documentclass[english,aps,prb,twocolumn,superscriptaddress,showpacs,floatfix]{revtex4}
\usepackage{amsmath}
\usepackage{graphicx}
\usepackage{amssymb}
\usepackage{color}

\makeatletter

\usepackage{babel}
\makeatother

\begin{document}
\newcommand{\bra}[1]{\left\langle #1\right|}
\newcommand{\ket}[1]{\left|#1\right\rangle }
\newcommand{\braket}[2]{\left\left \langle#1|#2\right\rangle }
\newcommand{\Tr}{\mathrm{Tr}}
\newcommand{\commute}[2]{\left[#1,#2\right]}
\newcommand{\anticommute}[2]{\left\{#1,#2\right\}}
\newcommand{\expect}[1]{\left\langle #1\right\rangle }
\newcommand{\sans}[1]{\mathsf{#1}}
\newcommand{\jcomment}[1]{\emph{\color[rgb]{0,0,0.6}(#1)}}
\newcommand{\bcomment}[1]{\emph{\color[rgb]{0.6,0,0}(#1)}}

\preprint{This line only printed with preprint option}

\title{Free-induction decay and envelope modulations in a narrowed nuclear spin bath}

\date{\today}

\author{W. A. Coish}
\affiliation{Institute for Quantum Computing and Department of Physics and Astronomy,
University of Waterloo, Waterloo, ON, N2L 3G1, Canada}
\affiliation{Kavli Institute for Theoretical Physics, UCSB, Santa Barbara, CA 93106, USA}

\author{Jan Fischer}
\affiliation{Department of Physics, University of Basel, Klingelbergstrasse 82,
4056 Basel, Switzerland}

\author{Daniel Loss}
\affiliation{Department of Physics, University of Basel, Klingelbergstrasse 82,
4056 Basel, Switzerland}
\affiliation{Kavli Institute for Theoretical Physics, UCSB, Santa Barbara, CA 93106, USA}

\begin{abstract}
We evaluate free-induction decay for the transverse components
of a localized electron spin coupled to a bath of nuclear spins via
the Fermi contact hyperfine interaction. Our perturbative treatment
is valid for special (narrowed) bath initial conditions and when the
Zeeman energy of the electron $b$ exceeds the total hyperfine coupling
constant $A$: $b>A$.  Using one unified and systematic method, we recover previous results reported at short and long times using different techniques.  We find a new and unexpected modulation of the free-induction-decay envelope, which is present even for a purely isotropic hyperfine interaction without spin echoes and for a single nuclear species.  We give sub-leading corrections to the decoherence rate, and show that, in general, the decoherence rate has a non-monotonic dependence on electron Zeeman splitting, leading to a pronounced maximum.  These results illustrate the limitations of methods that make use of leading-order effective Hamiltonians and re-exponentiation of short-time expansions for a strongly-interacting system with non-Markovian (history-dependent) dynamics.
\end{abstract}


\maketitle

\section{introduction}

The hyperfine interaction of a quantum-dot-confined electron spin with surrounding
nuclear spins and the resulting decoherence of electron-spin states
has been a focus of research in the last several years because of
potential applications in spintronics \cite{awschalom:book,
zutic:2004a, awschalom:2007a} and quantum information processing. \cite{loss:1998a,cerletti:2005a, hanson:2007a}
Hyperfine-induced electron-spin decoherence is one of the most significant obstacles to viable quantum computation with confined electron spins.  It is therefore of central importance to understand this decoherence mechanism so that schemes can be developed to suppress it.

One of the most promising strategies to suppress spin dephasing is to prepare the nuclear-spin
system in a less-noisy `narrowed' state. \cite{coish:2004a,klauser:2006a,stepanenko:2005a,giedke:2006a}  Once such a state is prepared, it can be maintained over an astonishingly long time scale, exceeding hours,\cite{greilich:2007a} since spin diffusion processes are highly suppressed near confined electron spins.\cite{klauser:2008a} Recently, great progress has been made in experimentally realizing such state
narrowing,\cite{greilich:2006a, greilich:2007a,reilly:2008a,greilich:2009a,latta:2009a,vink:2009a,xu:2009a} as well as single-spin readout and coherent control,\cite{petta:2005a,koppens:2006a,koppens:2007a,koppens:2008a,pioro-ladriere:2008a} which we expect to lead to improved coherence-time measurements in the very near future.

In the absence of refocusing pulses, and at time scales that are short compared to the relevant time scale for the nuclear dipolar interaction, an electron spin interacting with a narrowed nuclear-spin environment
dephases mainly via virtual flip-flops with the nuclear spins, provided
a sufficiently strong magnetic field is applied to suppress direct spin flips between electron and nuclear spins.
Under these conditions, the electron-spin dynamics pass through 
various stages with a zoo of different decay laws, obtained by various methods (see, e.g., Fig. 5 of Ref. \onlinecite{coish:2009a}): an exact solution for a fully-polarized nuclear system and leading-order generalized master equation (GME) have both shown
a short-time (partial) power-law decay, \cite{khaetskii:2002a, khaetskii:2003a, coish:2004a} an effective-Hamiltonian and short-time-expansion approach shows that the initial partial decay is followed by a quadratic decay shoulder, \cite{yao:2006a,liu:2007a} and a Born-Markov approximation applied to the same effective Hamiltonian shows that the majority of the decay is typically exponential in the high-field (perturbative) regime. \cite{coish:2008a, cywinski:2009a, cywinski:2009b}  Finally, an equation-of-motion approach has shown a long-time power-law decay to zero. \cite{deng:2006a,deng:2008a}

In this article, we show that each of these results can be obtained in a systematic way from a single unified approach, by extending the GME introduced in Ref. \onlinecite{coish:2004a} to higher order.  In addition to recovering previous results at all time scales, we find important qualitatively new features, including a modulation of the decay envelope (even for a fully isotropic hyperfine interaction).  Moreover, we give sub-leading corrections (in the inverse electron Zeeman splitting $1/b$) to the decoherence rate $1/T_2$ calculated previously.\cite{coish:2008a}  These corrections suggest an interesting non-monotonic dependence of $1/T_2$ on $b$.  Neither the envelope modulations, nor the sub-leading corrections to $1/T_2$ can be found from dynamics under the effective Hamiltonian alone. The results presented here therefore show limits to the validity of some previous approaches based on high-order expansions of a leading-order effective Hamiltonian.

The rest of this article is organized as follows: In Sec. \ref{sec:Hamiltonian} we introduce the relevant Hamiltonian, initial conditions, and exact equation of motion (GME) for the electron spin.  Sec. \ref{sec:self-energy} contains a review of the systematic expansion for the electron-spin self-energy (memory kernel) in powers of electron-nuclear flip-flops $V_\mathrm{ff}$, and gives the result up to fourth order in $V_\mathrm{ff}$.  In Sec. \ref{sec:spin-dynamics} we evaluate the full non-Markovian spin dynamics for an electron in a two-dimensional quantum dot. We recover previously-known results found using various other methods and present new results for the envelope modulation and corrections to the exponential decoherence rate $1/T_2$.  In Sec. \ref{sec:small-b}, we explore the behavior of the fourth-order solution in the non-perturbative regime and comment on the range of validity of this technique and possible extensions to higher order.  We conclude in Sec. \ref{sec:conclusions} with a summary of the results found here and a comparison of these results with those that have been presented in the literature.  Technical details are given in Appendixes \ref{sec:Self-energy-expansion} and \ref{sec:higher-order}.

\section{Hamiltonian and generalized master equation}\label{sec:Hamiltonian}

We consider the Hamiltonian for a localized electron spin-1/2 (with
associated spin operator $\mathbf{S}$), interacting with a bath of
nuclear spins $\mathbf{I}_{k}$ via the Fermi contact hyperfine interaction.
We allow generally for a Zeeman splitting $b=g\mu_{\mathrm{B}}B$
of the central spin $\mathbf{S}$ and site- (or species-) dependent
Zeeman splitting $b\gamma_{k}$ in the bath for a nuclear spin $\mathbf{I}_{k}$
at site $k$. The Hamiltonian for this system is (setting $\hbar=1$)
\begin{equation}
H=bS^{z}+b\sum_{k}\gamma_{k}I_{k}^{z}+\mathbf{S}\cdot\mathbf{h};\,\,\,\,\mathbf{h}=\sum_{k}A_{k}\mathbf{I}_{k},\label{eq:Hamiltonian}
\end{equation}
where the hyperfine coupling constant at site $k$ is given by $A_{k}=v_{0}A^{j_{k}}\left|\psi(\mathbf{r}_{k})\right|^{2}$
if the nucleus at site $k$ is of isotopic species $j_{k}$ with associated
total hyperfine coupling constant $A^{j_{k}}$, $\psi(\mathbf{r}_{k})$
is the electron envelope wavefunction, evaluated at site $\mathbf{r}_{k}$
(the position of the $k^{\mathrm{th}}$ nuclear spin), and $v_0$ is the
atomic volume.

In Eq. (\ref{eq:Hamiltonian}) we have neglected the nuclear dipole-dipole interaction, which can give rise to additional internal dynamics in the nuclear spin system, and consequent decay of the electron spin.\cite{klauder1962a,desousa:2003a,witzel:2006a,yao:2006a,cywinski:2009a}  Dipole-dipole-induced nuclear spin dynamics are highly suppressed in the presence of an inhomogeneous quadrupolar splitting\cite{maletinsky:2009a} or Knight-field gradient in a small quantum dot (the ``frozen-core'' or diffusion-barrier effect, see Ref. \onlinecite{ramanathan:2008a} for a review).  The relevant decoherence rate due purely to the hyperfine interaction is enhanced for a small dot ($1/T_2\sim 1/N$ for a quantum dot containing $N$ nuclear spins), whereas the dipole-dipole-induced nuclear dynamics are suppressed for a small dot due to the frozen-core effect.  Thus, there will always be some dot size $N$ where the nuclear dipolar interactions can be neglected, even up to times that are long compared to the electron-spin decoherence time.

\subsection{Initial conditions}

We choose product-state initial conditions\begin{equation}
\rho(0)=\rho_{S}(0)\otimes\rho_{I}(0),\label{eq:ProductStateIC}\end{equation}
where $\rho_{I(S)}=\mathrm{Tr}_{S(I)}\rho$ is the reduced density
matrix for the nuclear (electron) system. Such an initial state can
be prepared through fast strong pulses applied to the electron spin
or by allowing an electron to tunnel rapidly into a localized orbital.\citep{coish:2004a} 

Typically, nothing will be known about the nuclear-spin system at
the beginning of an experiment, and the density matrix $\rho_{I}(0)$
will be well-characterized by a completely random (infinite temperature)
mixture. Randomized initial conditions for the nuclear-spin bath result
in a rapid Gaussian decay of the transverse electron spin in the presence
of a strong Zeeman splitting $b$.\citep{khaetskii:2002a,merkulov:2002a,schliemann:2002a,schliemann:2003a,coish:2004a,petta:2005a}
This rapid decay, due to static fluctuations in the initial conditions,
can be removed by performing a measurement of the $z$-component of the slowly-varying
nuclear field $h^{z}$.\citep{coish:2004a} There have been several
theoretical proposals \citep{klauser:2006a,stepanenko:2005a,giedke:2006a}
to measure the nuclear-spin system into an eigenstate of $h^{z}$ and
there are now several experiments where similar state preparation has been
achieved through dynamical pumping.\citep{greilich:2007a,reilly:2008a,greilich:2009a,vink:2009a,latta:2009a,xu:2009a}
After preparing the nuclei in an eigenstate of the operator $h^{z}$,
the nuclear system will be described most generally by an arbitrary
mixture of degenerate $h^{z}$-eigenstates $\ket{n_{i}}$:
\begin{equation}
\rho_{I}(0)=\sum_{i}\rho_{ii}\ket{n_{i}}\bra{n_{i}}+\sum_{i\ne j}\rho_{ij}\ket{n_{i}}\bra{n_{j}},\label{eq:GeneralNuclearInitState}
\end{equation}
where 
\begin{equation}
h^{z}\ket{n_{i}}=h_{n}^{z}\ket{n_{i}}\quad\forall i.\label{eq:hzEigenvalueEquation}
\end{equation}
In this paper, we will assume that there is no `special' phase relationship
between the different $h^{z}$-eigenstates, which allows us to approximate
$\rho_{I}(0)$ by the diagonal part of Eq. (\ref{eq:GeneralNuclearInitState})
\begin{equation}\label{eq:RhoIZero}
\rho_{I}(0)\approx\sum_{i}\rho_{ii}\ket{n_{i}}\bra{n_{i}},
\end{equation}
 where, for any particular $i$, the state $\ket{n_{i}}$ is
given by a product of $I_{k}^{z}$-eigenstates: 
\begin{equation}
\ket{n_{i}}=\bigotimes_{k}\ket{I_{j_{k}}m_{k}^{i}},\label{eq:NuclearICs}
\end{equation}
with $I_{k}^{z}\ket{I_{j_{k}}m_{k}^{i}}=m_{k}^{i}\ket{I_{j_{k}}m_{k}^{i}}$ and where $-I_{j_k}\le m_k^i \le I_{j_k}$.

We will find it convenient to define the average of an arbitrary function of $I^z$-eigenvalues $f(m)$:
\begin{equation}\label{eq:avg-definition}
 \left<\left<f(m)\right>\right>=\sum_i\rho_{ii}\bra{n_i}f(m_k^i)\ket{n_i},
\end{equation}
where we assume a uniformly polarized nuclear-spin system throughout this article, making the average on the right-hand side independent of $k$.

\subsection{Generalized master equation}

The von Neumann equation for the density matrix $\dot{\rho}=-i\commute{H}{\rho}=-i\mathsf{L}\rho$ can be rewritten in the form of the Nakajima-Zwanzig generalized master equation.\citep{fick:1990a,breuerpetruccione} If we introduce a projection superoperator $\mathsf{P}$ that preserves the initial condition $\mathsf{P}\rho(0)=\rho(0)$, the Nakajima-Zwanzig GME can be written as 
\begin{eqnarray}
\mathsf{P}\dot{\rho}(t) & = & -i\mathsf{PLP}\rho(t)-i\int_{0}^{t}dt^{\prime}\mathsf{\Sigma}(t-t^{\prime})\mathsf{P}\rho(t^{\prime}),\label{eq:NZEquation}\\
\mathsf{\Sigma}(t) & = & -i\mathsf{PLQ}e^{-i\mathsf{LQ}t}\mathsf{QLP}.\label{eq:SelfEnergy}
\end{eqnarray}
Additionally, the projector $\mathsf{P}$
must satisfy $\sans{P}^{2}=\sans{P}$ and is typically chosen to preserve
all system variables $S^{\alpha}$: $\mathrm{Tr}S^{\alpha}\rho(t)=\mathrm{Tr}S^{\alpha}\mathsf{P}\rho(t)$.
Here, $\mathsf{Q}$ is the complement projector: $\mathsf{Q}=\mathsf{1-P}$.
For the special case of the Hamiltonian (Eq. (\ref{eq:Hamiltonian}))
and the initial condition (\ref{eq:RhoIZero}), we choose the projection
superoperator $\sans{P}=\rho_{I}(0)\mathrm{Tr}_{I}$ and find that
the exact equation of motion for the transverse electron spin $(S_{\pm}=S^{x}\pm iS^{y})$
is of the form\citep{coish:2004a}
\begin{eqnarray}
\frac{d}{dt}\left\langle S_{+}\right\rangle _{t} & = & i\omega_{n}\left\langle S_{+}\right\rangle _{t}-i\int_{0}^{t}dt^{\prime}\Sigma(t-t^{\prime})\left\langle S_{+}\right\rangle _{t^{\prime}},\label{eq:SplusEquation}\\
\Sigma(t) & = & \mathrm{Tr}\left[S_{+}\sans{\Sigma}(t)S_{-}\rho_{I}(0)\right],\label{eq:SigmaPlusPlusDef}\end{eqnarray}
where $\omega_{n}=b+h_{n}^{z}$.

\subsection{Rotating frame}

We define the coherence factor $x_t$ in the rotating frame
\begin{equation}\label{eq:xtRotatingFrame}
x_{t}=2e^{-i(\omega_{n}+\Delta\omega)t}\expect{S_{+}}_{t},
\end{equation}
and the associated self-energy
\begin{equation}
\label{eq:SigmaRotatingFrame}
\tilde{\Sigma}(t)=e^{-i(\omega_{n}+\Delta\omega)t}\Sigma(t),
\end{equation}
with Lamb shift $\Delta\omega$ due to virtual excitations of the bath:
\begin{equation}
\Delta\omega=-\mathrm{Re}\int_{0}^{\infty}dt\tilde{\Sigma}(t).\label{eq:DeltaOmegaDefinition}
\end{equation}
This gives an equation of motion for $x_{t}$:
\begin{equation}\label{eq:XtEOM1}
\dot{x}_{t}=-i\Delta\omega x_{t}-i\int_{0}^{t}dt^{\prime}\tilde{\Sigma}(t-t^{\prime})x_{t^{\prime}}.
\end{equation}
Eq. (\ref{eq:XtEOM1}) is an exact equation of motion for the coherence factor $x_t$, and therefore serves as an important starting point for sytematic approximations in the rest of this article.

\section{Self-energy expansion}\label{sec:self-energy}

In the absence of an exact closed-form expression for $\tilde{\Sigma}(t)$,
 we must resort to an approximation scheme. For a large electron-spin
Zeeman splitting $b$, and due to the large difference between the magnetic
moments of electron and nuclear spin ($\gamma_{k}\sim10^{-3}$), it
is appropriate to separate the Hamiltonian (Eq. (\ref{eq:Hamiltonian}))
into an unperturbed piece that preserves $S^{z}$ and a flip-flop
term, which induces energy non-conserving flip-flops between electron
and nuclear spins: $H=H_{0}+V_{\mathrm{ff}}$, where \begin{eqnarray}
H_{0} & = & \omega S^{z}+b\sum_{k}\gamma_{k}I_{k}^{z};\,\,\,\,\omega=b+h^{z},\label{eq:H0Definition}\\
V_{\mathrm{ff}} & = & \frac{1}{2}\left(h^{+}S_{-}+h^{-}S_{+}\right).\label{eq:VffDefinition}\end{eqnarray}
We can then write $\tilde{\Sigma}(t)$ in powers of $V_{\mathrm{ff}}$
by performing a Dyson-series expansion of Eq. (\ref{eq:SelfEnergy})
and inserting the result into the definition (Eq. (\ref{eq:SigmaPlusPlusDef})):\citep{coish:2004a}
\begin{equation}
\tilde{\Sigma}(t)=\tilde{\Sigma}^{(2)}(t)+\tilde{\Sigma}^{(4)}(t)+O\left(V_{\mathrm{ff}}^{6}\right).\label{eq:SelfEnergyExpansion}
\end{equation}
Progressively higher-order terms in the expansion involve a larger
number of flip-flops between the electron and nuclear bath spins.
Consequently, higher-order terms are suppressed by the energy cost
for such flip-flops, provided by the electron spin splitting $b$ for an unpolarized nuclear bath. In particular, up to factors of order unity and an overall common prefactor, the size of the $2(n+1)^{\mathrm{th}}$-order term is given by (see also Appendix A of Ref. \onlinecite{coish:2004a}): 
\begin{equation}\label{eq:HigherOrderSizeEstimate}
\tilde{\Sigma}^{(2[n+1])}(t)\propto\left(\frac{I(I+1)A}{b}\right)^{n}.
\end{equation}
For a nuclear spin of order unity ($I\sim1$), the condition for the
validity of a perturbative expansion in terms of $V_{\mathrm{ff}}$
(i.e., the condition for convergence of the series in Eq. (\ref{eq:SelfEnergyExpansion}))
is then given approximately by\citep{coish:2004a,deng:2006a,coish:2008a}\begin{equation}
b\gtrsim A.\label{eq:PerturbationCondition}\end{equation}

In the Born approximation, the self-energy $\tilde{\Sigma}(t)$ is
replaced by the leading-order non-vanishing term in the expansion
of Eq. (\ref{eq:SelfEnergyExpansion}): $\tilde{\Sigma}(t)\approx\tilde{\Sigma}^{(2)}(t)$.
To understand the evolution of $x_{t}$ within the Born approximation,
it is convenient to introduce the function $\psi(t)=\int_{t}^{\infty}dt^{\prime}\tilde{\Sigma}^{(2)}(t^{\prime})$,
which allows us to rewrite the equation of motion (Eq. (\ref{eq:XtEOM1}))
as\citep{fick:1990a}\begin{equation}
\dot{x}_{t}=-i(\psi(0)+\Delta\omega)x_{t}+\frac{d}{dt}R(t),\label{eq:MarkovNonMarkovSeparation}\end{equation}
where $R(t)=i\int_{0}^{t}dt^{\prime}\psi(t-t^{\prime})x_{t^{\prime}}$.
In a standard Born-Markov approximation, the dynamics induced by $R(t)$
are neglected. The real part of $\psi(0)$ cancels any remaining precession:
$-\mathrm{Re}\psi(0)=\Delta\omega$ (see Eq. (\ref{eq:DeltaOmegaDefinition}))
and the imaginary part of $\psi(0)$ gives rise to a purely exponential
decay of $x_{t}$ with decay rate $\Gamma=-\mathrm{Im}\psi(0)$. In
a previous calculation, it has been shown that the decay rate for
this system within a Born-Markov approximation vanishes: $\Gamma=0$.\citep{coish:2004a}
Within a Born approximation, all non-trivial dynamics in the rotating
frame are therefore induced by the non-Markovian remainder term $R(t)$.
The remainder term $R(t)$ has been investigated in detail previously\citep{khaetskii:2002a,khaetskii:2003a,coish:2004a}
and leads to a partial decay of the coherence factor of order\citep{coish:2004a}
$\left|R(t)\right|\sim O\left[\frac{1}{N}\left(\frac{A}{b}\right)^{2}\right]$
on a time scale $\sim N/A$, with long-time power-law tails. In the
rest of this article, we include the effects of the Born approximation
in inducing the Lamb shift (Eq. (\ref{eq:DeltaOmegaDefinition})),
but neglect the $\lesssim O(1/N)$ corrections due to $R(t)$ in the
perturbative regime. Additionally, we will include the fourth-order
correction to the self-energy in the expansion of Eq. (\ref{eq:SelfEnergyExpansion}),
which we show induces a more dramatic decay than the Born approximation,
albeit at a longer time scale. 

We now approximate the self-energy by including all terms at second
and fourth order in electron-nuclear spin flip-flops\begin{equation}
\tilde{\Sigma}(t)\approx\tilde{\Sigma}^{(2)}(t)+\tilde{\Sigma}^{(4)}(t).\label{eq:SelfEnergyBornAndRest}\end{equation}
Inserting Eq. (\ref{eq:SelfEnergyBornAndRest}) into Eq. (\ref{eq:XtEOM1})
we find, neglecting the dynamics with amplitude suppressed by $\sim1/N$
in the perturbative regime due to $R(t)$:
\begin{eqnarray}
\dot{x}_{t} & = & -i\int_{0}^{t}dt^{\prime}\tilde{\Sigma}^{(4)}(t-t^{\prime})x_{t^{\prime}},\label{eq:XtEOM2}\\
\Delta\omega & \approx & -\mathrm{Re}\int_{0}^{\infty}dt\tilde{\Sigma}^{(2)}(t).\label{eq:DeltaOmegaApprox}
\end{eqnarray}

The integro-differential equation (Eq. (\ref{eq:XtEOM2})) is difficult
to solve, in general. However, in terms of Laplace-transformed variables,
this equation becomes an algebraic equation, which can be solved directly.
Introducing the Laplace transform of some function $f(t)$,
\begin{equation}
f(s)=\int_{0}^{\infty} dt \, e^{-st}f(t), \quad \mathrm{Re}(s)>0, \label{xsdef}
\end{equation}
we rewrite Eqs. (\ref{eq:XtEOM2}) and (\ref{eq:DeltaOmegaApprox})
as
\begin{eqnarray}
x(s) & = & \frac{x_{0}}{s+i\tilde{\Sigma}^{(4)}(s)},\label{eq:coherence_laplace} \\
\Delta\omega & \approx & -\mathrm{Re}\tilde{\Sigma}^{(2)}(s=0^{+}).
\end{eqnarray}
 We have calculated the self-energy $\tilde{\Sigma}$ in Appendix \ref{sec:Self-energy-expansion}, including terms up to fourth order in $V_\mathrm{ff}$.  

\begin{figure}
\includegraphics[width=0.45\textwidth]{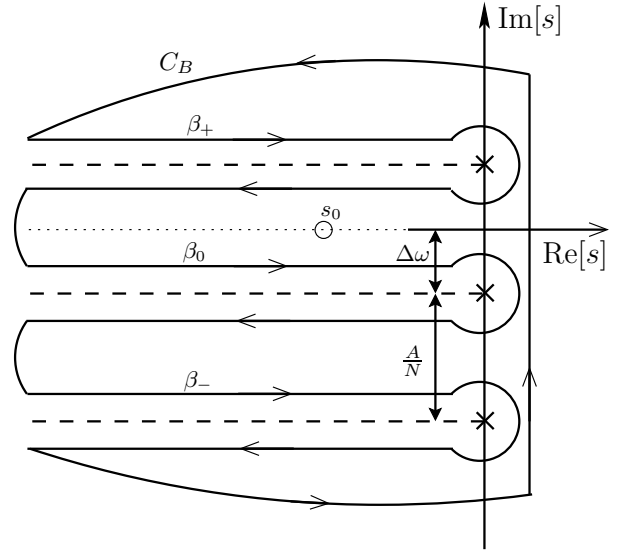}
\caption{\label{fig:Contour}(Color online) Contour used to evaluate the Bromwich inversion integral (in the rotating frame defined by Eq. (\ref{eq:xtRotatingFrame})).
The dynamics of the coherence factor $x_t$ are determined by a single pole at $s_0=-1/T_2$
and three branch cuts (see main text).  The pole is offset from the branch cuts by the Lamb shift $\Delta\omega$ and the excitation bandwidth (separation between branch points) is given by the size of the hyperfine coupling to a single nucleus at the center of the quantum dot, $A/N$.}
\end{figure}

The dominant contributions to $\tilde{\Sigma}(s)$ occur for $|s|\ll|\omega_n|$ in the rotating frame (high-frequency, $s\simeq i\omega_n$ in the lab frame).  We have expanded the second- and fourth-order self-energies in this limit, as described in Appendix A.  Corrections to this expansion are smaller than the retained contributions by a factor of $A/N\omega_n\ll 1$.  For explicit calculation, it is useful to specialize to the case of a homonuclear system (where $\gamma_k=\gamma$ for all $k$) and a two-dimensional quantum dot with Gaussian envelope function, leading to coupling constants\cite{coish:2004a,coish:2008a} $A_k=(A/N)e^{-k/N}$. Performing the continuum limit ($\sum_k\to\int dk$) and evaluating the relevant energy integrals for a uniformly polarized nuclear spin system leads to 
\begin{eqnarray}
&\tilde{\Sigma}^{(4)}(s-i\Delta\omega) = i\alpha\left[F_{+}(s)J_{+}(s)+F_{-}(s)J_{-}(s)-s\right], \label{eq:sigma4}\\
&\Delta\omega \simeq -\tilde{\Sigma}^{(2)}(s=0^+)  =  \frac{1}{8}(c_{+}+c_{-})\frac{A}{\omega_n}\frac{A}{N},
\end{eqnarray}
with 
\begin{equation}
   \alpha = \frac{c_+ c_-}{24}\left(\frac{A}{\omega_n}\right)^2,
\end{equation}
and where $c_\pm = I(I+1)- \langle\langle m(m \pm 1) \rangle\rangle$ are the coefficients introduced in Ref. \onlinecite{coish:2004a} with $\langle\langle\cdots\rangle\rangle$ indicating an average over the mixture of $I_k^z$-eigenvalues, described by Eq. (\ref{eq:avg-definition}).  Additionally, we have introduced the functions
\begin{eqnarray}
F_{\pm}(s) & = & \left(\frac{N}{A}\right)^2\left(s\pm i\frac{A}{N}\right)^{2}\left(s\mp2i\frac{A}{N}\right),\\
J_{\pm}(s) & = & \log\left(s\pm i\frac{A}{N}\right)-\log(s).
\end{eqnarray}

After inserting Eq. \eqref{eq:sigma4} into Eq. \eqref{eq:coherence_laplace}, we find that the Laplace-transformed coherence factor has three branch points, and if the principle branch is chosen for all branch cuts and at large electron Zeeman splitting ($b\gg A$), there is one pole at $s=s_0$ (see Fig. \ref{fig:Contour}).  These non-analytic features determine the dynamics of the coherence factor (see below).  The equation-of-motion method adopted in Refs. \onlinecite{deng:2006a,deng:2008a} bears some similarity to the current approach.  However, the excitation bandwidth (distance between branch points) found in Refs. \onlinecite{deng:2006a,deng:2008a} is half that found here ($A/2N$ rather than $A/N$), leading to a difference (by a factor of 2) for relevant decay time scales. We comment on other differences, below.

\begin{figure}
\includegraphics[scale=0.8]{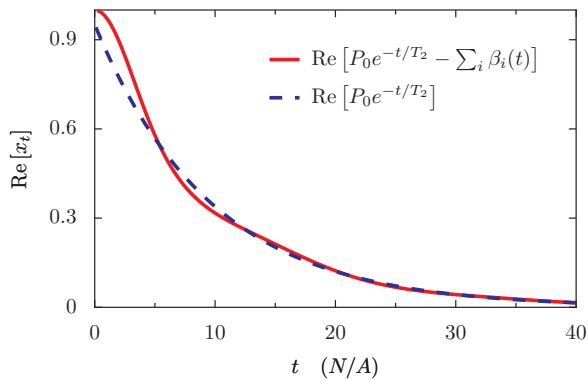}
\caption{\label{fig:Markov_comparison}(Color online) Comparison of the contribution from the pole at $s_{0}$ in Fig. \ref{fig:Contour} with exponentially decaying residue
(blue dashed line) with the full fourth-order result, obtained numerically (red solid line). We take the initial condition $x_0=1$, assume an unpolarized nuclear spin system [with $\omega_n=b$, $c_\pm = \frac{2}{3}I(I+1)$, which follows from $\left<\left<m^2\right>\right>=I(I+1)/3$ if all Zeeman levels have equal population], and have chosen $I=3/2$ and $A/b=1/3$.}
\end{figure}
\begin{figure}
\includegraphics[scale=0.8]{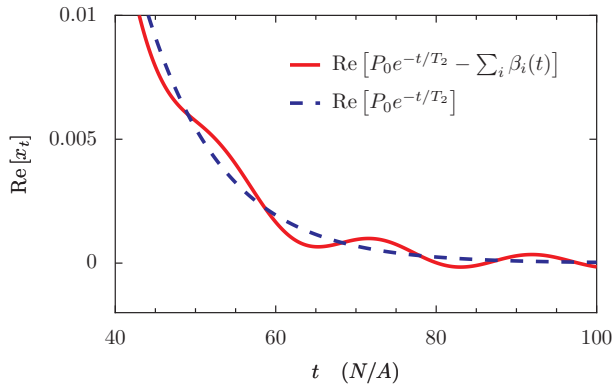}
\caption{\label{fig:longtime_decay}(Color online) Long-time decay. At long times, the exponential decay envelope is modulated by branch-cut contributions.  Parameters are as in Fig. \ref{fig:Markov_comparison}.}
\end{figure}

\section{Spin dynamics}\label{sec:spin-dynamics}

We find the time-dependent coherence factor by evaluating the Bromwich inversion integral
\begin{equation}
 x_t  =  \lim_{\gamma\to 0^+}\frac{1}{2\pi i}\int_{\gamma-i\infty}^{\gamma+i\infty}ds e^{st}x(s),
\end{equation}
which can be rewritten in terms of an integral over the closed contour $C_B$ and branch-cut integrals $\beta_j$, $j=0,+,-$ (see Fig. \ref{fig:Contour}):
\begin{eqnarray}\label{eq:xtintegral}
x_t & = & \frac{1}{2\pi i}\oint_{C_B} ds e^{st}x(s)  -\sum_j \beta_j(t)\nonumber\\
   & = & \mathrm{Res}\left[e^{st} x(s),s=s_0\right]-\sum_j \beta_j(t).\label{eq:xtintegral-residue}
\end{eqnarray}
In Eq. \eqref{eq:xtintegral-residue}, we have applied the residue theorem to write the integral over $C_B$ in terms of a residue at the pole $s_0$.

Since the rotating frame is chosen to give $s_0=-1/T_2$ purely real, we find the general result
\begin{equation}
x_{t}=P_0 e^{-t/T_2}-\sum_{i}\beta_{i}(t),
\end{equation}
with $P_0$ given by Eq. (\ref{eq:P0Residue}), below.  The coherence factor is characterized by two terms: an exponential, which dominates in the perturbative regime ($A\lesssim b$, see Fig. \ref{fig:Markov_comparison}), and a sum of branch-cut integrals, which give rise to modulations of the decay envelope and a dominant long-time power-law decay (see Fig. \ref{fig:longtime_decay}).

\subsection{Envelope modulations and long-time decay}

\begin{figure}
\includegraphics[scale=0.8]{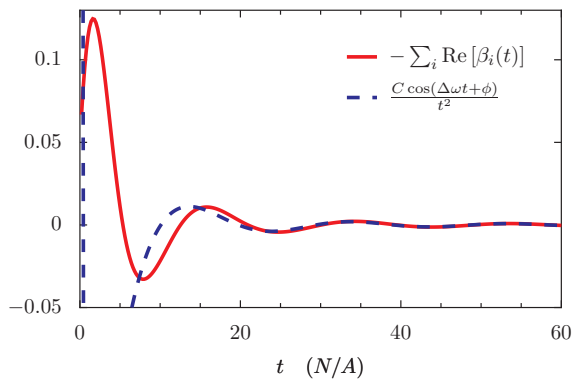}
\caption{\label{fig:branchcut_comparison}(Color online) Comparison of the full numerical branch-cut integral $-\sum_i\mathrm{Re}\left[\beta_i(t)\right]$ with
the long-time asymptotic expression given by Eq. \eqref{eq:xt-asymptotic}. Parameters are as in Fig. \ref{fig:Markov_comparison}.  }
\end{figure}

From direct asymptotic analysis of the branch-cut integrals,
we find $\beta_\pm(t) \propto 1/t^3$, while $\beta_0(t) \propto 1/t^2$ at long times.  Since the pole contribution decays exponentially, the leading long-time asymptotics of $x_t$ are thus given by $\beta_0(t)$. Evaluating the prefactor, we find the long-time limit  (valid for $t\gg \mathrm{max}\left[1/\Delta\omega,\left(6\alpha/\Delta\omega\right)\ln\left|N\Delta\omega/6\alpha A\right|\right]$):
\begin{equation}
	\beta_0(t) \sim -\frac{6 \alpha x_0}{(2\pi\alpha A/N-i \Delta \omega)^2}
	\, \frac{e^{-i \Delta \omega t}}{t^2},\label{eq:branchcut_integral}
\end{equation}
which gives the long-time behavior of the coherence factor with initial condition $x_0=1$ (see Fig. \ref{fig:branchcut_comparison}):
\begin{equation}\label{eq:xt-asymptotic}
 \mathrm{Re}[x_t] \sim \frac{C \cos\left(\Delta\omega t+\phi\right)}{t^2},
\end{equation}
where
\begin{eqnarray}
 C &=& \frac{6\alpha}{\left(2\pi \alpha A/N\right)^2+\Delta\omega^2},\\
\phi & = & -2 \arctan \left(\frac{\Delta\omega N}{2\pi \alpha A}\right).
\end{eqnarray}

The modulations at a frequency $\Delta\omega$ in Eq. (\ref{eq:xt-asymptotic}) can be understood on physical grounds in the following way: The short-time dynamics of the electron spin are controlled by nuclear spins near the center of the dot, which are coupled most strongly.  The effective precession frequency of the electron spin is therefore renormalized by the shift $\Delta\omega$ due to virtual flip-flop processes with nuclei near the center for most of the decay envelope.  The long-time decay, however, is controlled by weakly-coupled nuclei far from the center of the dot, which cannot strongly shift the electron-spin precession frequency.  The long-time dynamics therefore occur at the `bare' precession frequency $\omega_n$.  The difference in frequency between the dominant (short-time) and sub-dominant (long-time) behavior leads to a relative beating at the frequency difference $\Delta\omega$. We note that the physical origin of this envelope modulation is completely different from the more typical case of electron spin-echo envelope modulation (ESEEM), which is often observed in systems with an anisotropic hyperfine interaction.\cite{rowan:1965a,mims:1972a,childress:2006a,witzel:2007b}  The modulations described here occur even in the present case of a purely isotropic interaction and without spin echoes.  The two cases can be experimentally distinguished through a difference in the magnetic-field dependence of the modulation frequency.  Finally, we note that the modulations found here are reminiscent of modulations in branch-cut contributions that have been highlighted previously for the spin-boson model.\cite{divincenzo:2005a}

Another striking feature of Eq. (\ref{eq:xt-asymptotic}) is the long-time power-law tail.  This differs from the long-time exponential decay found by other authors\cite{liu:2007a,cywinski:2009a,cywinski:2009b} using re-summation and re-exponentiation techniques.
The same long-time power-law decay ($\propto 1/t^2$) has previously been predicted in Refs. \onlinecite{deng:2006a,deng:2008a} based on an equation-of-motion method, but without mention of the phase shift or envelope modulations predicted by Eq. \eqref{eq:xt-asymptotic}. To compare the results given here directly with Refs. \onlinecite{deng:2006a,deng:2008a}, we consider the case of $I=1/2$ with an unpolarized nuclear-spin bath.  This gives
\begin{eqnarray}
 C & = & \frac{4}{(A/N)^2}+\mathcal{O}\left[\left(\frac{A}{b}\right)^2\right],\\
 \phi & = & -\pi +\mathcal{O}\left[\left(\frac{A}{b}\right)^3\right].
\end{eqnarray}
Although the power law found here matches that reported in Refs. \onlinecite{deng:2006a,deng:2008a}, and the modulations or phase shift can be ignored in the limit $A/b\ll 1$, the prefactor $C$ (which is\footnote{$A/b$ in the units used here is equivalent to $N/\Omega$ in the units of Refs. \onlinecite{deng:2006a,deng:2008a}.} $C=\mathcal{O}\left[(b/A)^2\right]$ in Refs. \onlinecite{deng:2006a,deng:2008a}) is qualitatively different.  In particular, here we find that the power-law contribution with modulations can have substantial weight (of order unity) in the perturbative regime $A\lesssim b$.  This is clear from Fig. \ref{fig:branchcut_comparison}, where we show that the branch-cut contributions can contribute approximately 10\% of the total decay amplitude.

\subsection{Decay shoulder}
\begin{figure}
\includegraphics[scale=0.8]{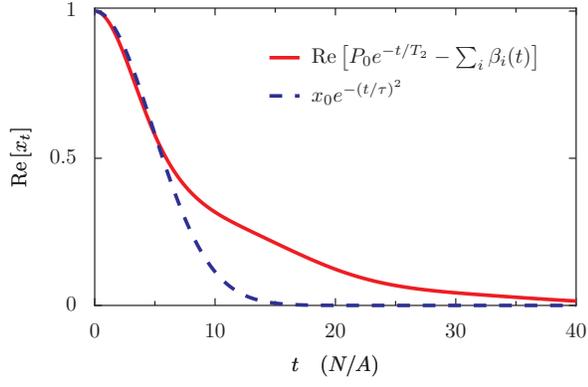}
\caption{\label{fig:initial_decay}(Color online) Decay shoulder. Here we show a comparison of the short-time quadratic decay with the full non-Markovian envelope for $I=3/2$ and $A/b=1/3$.}
\end{figure}

For small $t$, we perform a Taylor-series expansion of $x_t$:
\begin{equation}
	x_t = x_0 + \dot{x}_0 t + \frac{1}{2} \, \ddot{x}_0 t^2 + \ldots.
\end{equation}
From Eq. \eqref{eq:XtEOM2} and the initial value theorem we find $\dot{x}_0 = 0$ and $\ddot{x}_0=-i\tilde{\Sigma}^{(4)}(t=0)=-i\lim_{s\to\infty}s\tilde{\Sigma}^{(4)}(s)$.  Inserting Eq. \eqref{eq:self-energy-discrete} for $\tilde{\Sigma}^{(4)}(s)$ and choosing $x_0=1$ gives:
\begin{equation}\label{eq:decay-shoulder}
	x_t \simeq 1 - \frac{t^2}{\tau^2}, \quad
	\tau \simeq \sqrt{\frac{2 \omega_n^2}{c_+ c_- \left(\sum_k A_k^2\right)^2}}.
\end{equation}
Eq. \eqref{eq:decay-shoulder} gives the same short-time decay reported in Ref. \onlinecite{liu:2007a}, which was taken to describe a Gaussian coherence decay: $x_t\simeq x_0\exp{\left[-(t/T_{2,A})^2\right]}$.  Here we note that the Gaussian approximation is only valid for times less than the actual decay time ($t\ll \tau$)  in the perturbative regime ($b\gtrsim A$) since the dominant decay is exponential in this regime for a typical (two-dimensional parabolic) quantum dot, as emphasized in Ref. \onlinecite{coish:2008a}, and illustrated here in Figs. \ref{fig:Markov_comparison} and \ref{fig:initial_decay}.  Re-exponentiation also fails for the fourth-order solution at lower magnetic field, as we show in Sec. \ref{sec:small-b}, below.

For a uniform unpolarized nuclear spin system, and for an electron with Gaussian envelope function in two dimensions, we find
\begin{equation}
	\tau \simeq \frac{6 \sqrt{2}}{I(I+1)} \left( \frac{b}{A} \right)\left( \frac{N}{A} \right).
\end{equation}
We compare the initial decay found using this formula with the full non-Markovian solution in Fig. \ref{fig:initial_decay}.  While the short-time decay shoulder is well-described by a Gaussian, the full decay envelope is much better described by the dominant exponential solution (see Fig. \ref{fig:Markov_comparison}).  At larger Zeeman splitting $b$, the distinction between Gaussian and exponential becomes significantly more pronounced.

\subsection{Exponential decay}\label{sec:exponential-decay}
We evaluate the residue at the pole $s_0$ in Fig. \ref{fig:Contour}, giving
\begin{equation}\label{eq:P0Residue}
 P_0 =\frac{1}{1+i\left.\frac{d}{ds}\tilde{\Sigma}^{(4)}(s)\right|_{s=-1/T_2}}. 
\end{equation}
For a two-dimensional parabolic quantum dot, with an unpolarized nuclear system, we find $P_0= 1+\mathcal{O}\left(\left[\frac{A}{b}\right]^2\ln\left[\frac{A}{b}\right]\right)$.  Thus, when $A<b$, a Markov approximation is justified (resulting in a dominant exponential decay), in agreement with the conclusions of Refs. \onlinecite{coish:2008a,cywinski:2009a,cywinski:2009b}.

In the Markovian regime, the decay rate for the exponentially decaying pole can be determined through\cite{coish:2008a} $1/T_2=-\mathrm{Im}\tilde{\Sigma}^{(4)}(0^+)$.  From the self-energy given in Eq. (\ref{eq:self-energy-discrete}), this gives 
\begin{equation}\label{eq:homonucT2}
 \frac{1}{T_2} = \frac{\pi c_+ c_-}{4\omega_n^2}\sum_{k,k'} A_k^2 A_{k'}^2\delta(A_k-A_{k'}-\Delta\omega).
\end{equation}
Here, the Markovian decay rate depends on the density of states for pair flips at an energy determined by the Lamb shift $\Delta\omega\propto A/\omega_n$.  The presence of $\Delta\omega$ in the energy-conservation condition can be understood physically as arising from the rapid initialization that we assumed, giving rise to the product-state initial condition (Eq. (\ref{eq:ProductStateIC})).  At the instant the flip-flop interaction $V_\mathrm{ff}$ is `turned on', the electron spin experiences only the bare precession frequency $\omega_n$.  However, after some interaction time scale, the renormalized precession frequency $\omega_n+\Delta\omega$ gives the electron energy splitting, and so the correct energy-conservation condition for nuclear-spin pair flips contains the difference of these two quantities (i.e., $\Delta\omega$). The dependence on $\Delta\omega$ shown in Eq. (\ref{eq:homonucT2}) results in an interesting (in general, non-monotonic) dependence of $1/T_2$ on magnetic field.\footnote{This non-monotonic dependence of $1/T_2$ on the electron-spin splitting is reminiscent of a similar effect found in the spin-boson model.  There, a non-monotonic dependence of the decoherence rate as a function of energy splitting is found when properly accounting for renormalization factors.\cite{divincenzo:2005a}}  In contrast, the effective-Hamiltonian approach that was adopted in Refs. \onlinecite{yao:2006a,liu:2007a,coish:2008a,cywinski:2009a,cywinski:2009b} incorporates the Lamb shift as an additive constant directly into the electron-spin Zeeman splitting, and so it does not enter into the formula for $1/T_2$.  The Lamb shift that comes out of the same procedure used here, but starting from the effective Hamiltonian,\cite{coish:2008a} has a dependence $\Delta\omega\propto (A/\omega_n)^2$, so the effective-Hamiltonian treatment does not recover the correct magnetic-field dependence given by Eq. \eqref{eq:homonucT2}, although the leading-order $\Delta\omega\simeq 0$ behavior is recovered correctly.  This is not surprising, since the effective Hamiltonian is only strictly valid to leading order in $A/\omega_n$.  Through explicit calculation at higher orders, we have checked that the leading correction to the Markovian decay rate at sixth order in $V_\mathrm{ff}$ is $\mathcal{O}\left[\left(A/\omega_n\right)^4\right]$ (see Appendix \ref{sec:higher-order}), and so the expression given here is at least correct up to and including terms of order $\mathcal{O}\left[\left(A/\omega_n\right)^3\right]$.  

\begin{figure}
\includegraphics[scale=0.8]{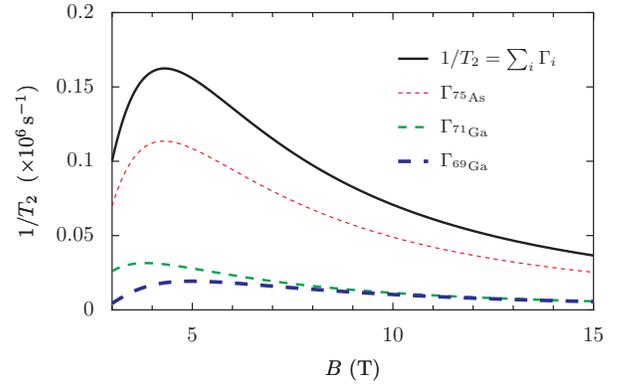}
\caption{\label{fig:gaasrates}(Color online) Decoherence rates $\Gamma_{i}$ from Eq. (\ref{eq:Gamma-i})
and total decoherence rate $1/T_2=\sum_{i}\Gamma_{i}$ for an electron
spin in a GaAs quantum dot containing $N=10^{5}$ nuclei with $g$-factor $|g|=0.4$. The decoherence rate shows a non-monotonic behavior, reaching a pronounced maximum.  This is in contrast to the leading-order result,\cite{coish:2008a} and in contrast to the results from other higher-order expansions.\cite{cywinski:2009b} We have used hyperfine coupling constants $A^{^{69}Ga}=74\,\mu e\mathrm{V}$, $A^{^{71}Ga}=96\,\mu e\mathrm{V}$, $A^{^{75}As}=86\,\mu e\mathrm{V}$ and relative abundances $\nu_{^{69}Ga}=0.3$, $\nu_{^{71}Ga}=0.2$, $\nu_{^{75}As}=0.5$, which have been estimated in Ref. \onlinecite{paget:1977a} (see Table 1 of Ref. \onlinecite{coish:2009a}).}
\end{figure}

Specializing to a two-dimensional quantum dot with Gaussian envelope function and evaluating the energy integrals in the continuum limit gives
\begin{equation}\label{eq:T2-2d-dot}
 \frac{1}{T_2} = \frac{8\pi c_+ c_-}{3(c_+ + c_-)^2}\frac{A}{N} (\epsilon^5-3\epsilon^3+2\epsilon^2)\Theta(1-\epsilon),
\end{equation}
where
\begin{equation}
 \epsilon = \frac{c_+ + c_-}{8}\left|\frac{A}{\omega_n}\right|.
\end{equation}
First, we note that the sub-leading contribution to $1/T_2$ in Eq. \eqref{eq:T2-2d-dot} ($\propto \epsilon^3$) is suppressed only by one power of $A/\omega_n$ (up to corrections of order unity).  Second, this sub-leading correction has the opposite sign of the leading ($\sim\epsilon^2$) term, potentially leading to a non-monotonic dependence of $1/T_2$ on the electron Zeeman splitting when $\epsilon\sim 1$.  This non-monotonic dependence can be understood in the following way: As the electron Zeeman energy decreases from a large value $b\gg A$, the perturbative Lamb shift $\Delta\omega\propto 1/b$ increases, eventually reaching the edge of the band of single nuclear pair-flip excitations when $\Delta\omega\sim A/N$, at which point there are no more energy-conserving flip-flop processes.  For still lower magnetic fields, higher-order processes are required to conserve energy, but we find that these processes are further suppressed by the small parameter $c_+c_-$  for a polarized nuclear spin system ($c_+c_-\propto(1-p^2)$ for nuclear spin $I=1/2$) (see Appendix \ref{sec:higher-order}). The qualitative non-monotonic magnetic-field dependence described by Eq. (\ref{eq:T2-2d-dot}) will therefore apply at least in the case of a polarized nuclear spin system, even when higher-order terms in $V_\mathrm{ff}$ are taken into account.

\begin{figure}
\includegraphics[scale=0.8]{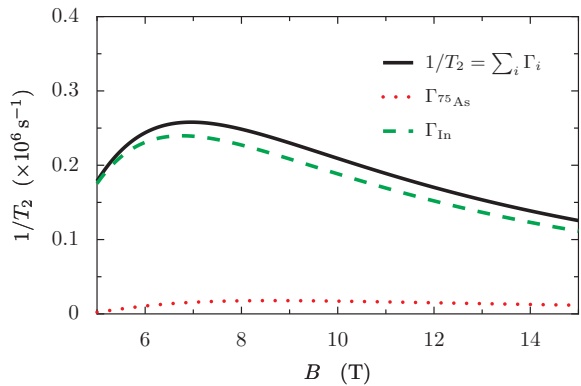}
\caption{\label{fig:ingaas-rates}(Color online) Total decoherence rate $1/T_2=\sum_{i}\Gamma_{i}$ (solid line) with $\Gamma_{i}$ from Eq. (\ref{eq:Gamma-i}) for an electron
spin in a $\mathrm{In}_x \mathrm{Ga}_{1-x} \mathrm{As}$ quantum dot containing 
$N=10^{5}$ nuclei with $g$-factor $|g|=0.5$ and In doping $x=0.3$.  We show individual contributions from $\Gamma_{^{75}\mathrm{As}}$ (dotted) and $\Gamma_\mathrm{In}$ (dashed).  Contributions from the gallium isotopes are not visible on this scale. For this plot, we have taken hyperfine coupling constants for the gallium and arsenic isotopes as in Fig. \ref{fig:gaasrates} and have used $A^{^{113}\mathrm{In}}=A^{^{115}\mathrm{In}}=A^{\mathrm{In}}=110\,\mu e\mathrm{V}$ from Ref. \onlinecite{liu:2007a} (see Table 1 of Ref. \onlinecite{coish:2009a}). }
\end{figure}

\begin{figure}
\includegraphics[scale=0.8]{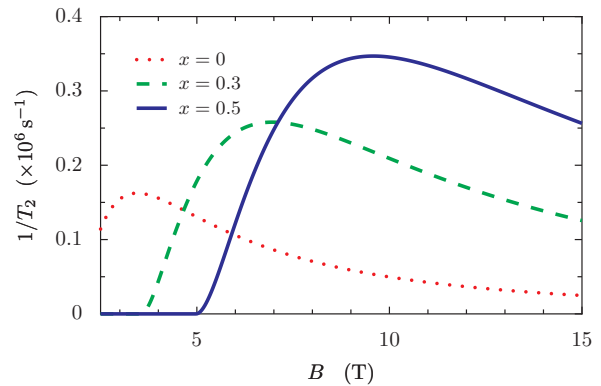}
\caption{\label{fig:ingaas-doping}(Color online) Decoherence rate $1/T_2$ for an electron spin in a $\mathrm{In}_x \mathrm{Ga}_{1-x} 
\mathrm{As}$ quantum dot containing $N=10^{5}$ nuclei with $g$-factor $|g|=0.5$ and In doping of $x=0$, $x=0.3$, and $x=0.5$.  Hyperfine coupling constants are as given in the caption of Fig. \ref{fig:ingaas-rates}.}
\end{figure}

It is straightforward to extend the analysis of this section to the case of a heteronuclear system.  Provided the difference in nuclear Zeeman energies exceeds the excitation bandwidth ($|(\gamma_i-\gamma_j)b|>A/N$), the decoherence rate is given by a sum of contributions from each nuclear species $i$: $1/T_2 = \sum_i \Gamma_i$, where, for a two-dimensional quantum dot with Gaussian envelope function, we find
\begin{eqnarray}
	\Gamma_{i} & = & \pi\nu_i^2\alpha_i\left(\epsilon_i^3-3\epsilon_i+2\right)\frac{A^i}{N}\Theta\left(1-\epsilon_i\right),\label{eq:Gamma-i}\\
	\epsilon_i & = & \left|\frac{N\Delta\omega}{A^{i}}\right|,\quad \Delta\omega = \sum_i \nu_i \frac{\left(c_+^i +c_-^i\right)}{8\omega_n}\frac{A^i}{\omega_n}\frac{A^i}{N},
\end{eqnarray}
and where we have introduced
\begin{equation}
 \alpha_i = \frac{c_{+}^{i}c_{-}^{i}}{24}\left(\frac{A^i}{\omega_n}\right)^2,
\end{equation}
with coefficients $c_{\pm}^i=I^i(I^i+1)-\left<\left<m^i(m^i\pm 1)\right>\right>$ for each isotopic species $i$.  We show the magnetic-field dependence of the decoherence rate $1/T_2$ from Eq. \eqref{eq:Gamma-i} in Fig. \ref{fig:gaasrates} for a GaAs quantum dot, and in Fig. \ref{fig:ingaas-rates} for an InGaAs quantum dot with a typical indium doping of $x=0.3$.  The dependence of the $1/T_2$ curve on indium doping $x$ for an InGaAs quantum dot is illustrated in Fig. \ref{fig:ingaas-doping}, where we see that the position of the maximum in the $1/T_2$ curve depends strongly on the concentration of the large-spin isotope (indium).  An experimental confirmation of this dependence of the maximum in $1/T_2$ as a function of indium doping would be a strong confirmation of this theory.

\section{Non-perturbative regime: $b\lesssim A$}\label{sec:small-b}

Although the expression we have given for the self-energy is strictly valid only in the perturbative regime ($b\gg A$), here we explore the non-Markovian dynamics of this solution outside of the regime of strict validity and comment on where the results become unphysical.

\begin{figure}
\includegraphics[scale=1]{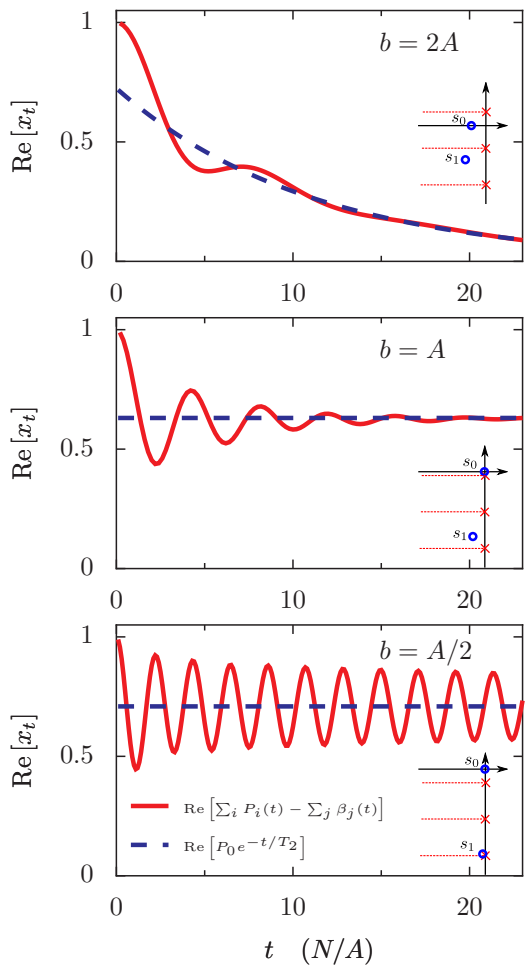}
\caption{\label{fig:b_dependence}(Color online) Decay envelopes calculated from numerical evaluation of branch-cut integrals and pole contributions for a two-dimensional quantum dot with Gaussian envelope function (solid lines).  Dashed lines show the contributions from the dominant pole at $s=s_0$. When the electron Zeeman splitting $b$ is below some critical value $b<b_c\sim A$, a second exponentially-decaying pole appears, leading to a biexponential decay with strong envelope modulations (top, $b=2A$).  When $b$ is decreased further, the dominant pole moves to the real axis (middle, $b=A$).  At still smaller values of $b$, the sub-dominant exponential pole has a vanishing decay rate (bottom, $b=A/2$), leading to sustained oscillations.  All plots correspond to an unpolarized narrowed nuclear bath with $I=3/2$.  Insets illustrate the approximate relative positions of poles (circles) and branch points (crosses) in each case.}
\end{figure}

We find the positions of poles and evaluate residues and branch-cut integrals numerically to find the coherence factor in this regime.  We consider the case of an unpolarized homonuclear spin system with spin $I=3/2$, appropriate to GaAs.  As the electron Zeeman splitting $b$ is lowered from $b\gg A$, we find there is a critical value of $b$ (near $b\simeq 2A$), below which there is a second pole (at $s=s_1$) with exponentially-decaying residue.  The coherence factor $x_t$ is then given by a sum over two pole contributions and three branch-cut integrals:
\begin{equation}
 x_t=\sum_{i=0,1} P_i(t)-\sum_{j=0,+,-} \beta_j(t).
\end{equation}
For $b=2A$ (top panel of Fig. \ref{fig:b_dependence}), there are two exponentially-decaying pole contributions, giving rise to a bi-exponenatial decay with strong envelope modulations corresponding to the difference in the imaginary part of the two poles.  At smaller Zeeman energy $A/2 \lesssim b \lesssim A$ (e.g. $b=A$ in the center panel of Fig. \ref{fig:b_dependence}), the pole at $s=s_0$ leaves the continuum band and merges with the imaginary axis, leading to a constant contribution $P_0(t)=P_0$, independent of $t$.  For still lower Zeeman splitting $b\lesssim A/2$ ($b=A/2$ in the lower panel of Fig. \ref{fig:b_dependence}), the second pole at $s=s_1$ leaves the continuum band at lower frequency and also merges with the imaginary axis.  In this regime, the only decay in the fourth-order solution is due to the small contribution from branch cuts, although envelope modulations remain.

The effects in the two lower panels of Fig. \ref{fig:b_dependence} demonstrate the danger of re-exponentiation of short-time behavior for a system where strong non-Markovian (history-dependent) effects become important.  The non-decaying fractions shown in Fig. \ref{fig:b_dependence} are, however, unphysical consequences of extending the solution to a regime of electron Zeeman splitting where it does not apply.  We expect higher-order corrections to the self-energy to broaden the continuum band as higher-order nuclear pair flips are included, resulting in several exponential decay time scales as the electron Zeeman energy is lowered.  Nevertheless, we have found that processes that can broaden the continuum band will be suppressed even at small electron Zeeman splitting $b\lesssim A$, provided the nuclear-spin system is polarized (see Appendix \ref{sec:higher-order}), and so some of this behavior will survive at least for a polarized nuclear-spin environment.  Whether perturbation theory can be controlled at \emph{any} polarization for $b<A$ through an adequate resummation of relevant terms, as suggested in Refs. \onlinecite{cywinski:2009a,cywinski:2009b}, is still unclear with the present method.

\section{Conclusions}\label{sec:conclusions}

We have investigated transverse-spin dynamics for an electron confined to a quantum dot, interacting with a bath of nuclear spins via the Fermi contact hyperfine interaction.  Using one unified technique, we have recovered results that have previously been reported using several different methods. These results include an initial partial decay, followed by a quadratic shoulder, a dominant exponential decay, and a long-time power-law tail.  Our results for the long-time behavior differ from those of Refs. \onlinecite{yao:2006a,liu:2007a,cywinski:2009a,cywinski:2009b}.  Here, we have found a long-time power-law decay ($\sim 1/t^2$), in contrast to the long-time exponential decay found by those authors.  While the decay law $\sim 1/t^2$ matches that found previously using an equation-of-motion approach,\cite{deng:2006a,deng:2008a} the prefactor found in the present work has a qualitatively different dependence on magnetic field.  In contrast to earlier works, which argue in favor of a regime of Gaussian decay,\cite{yao:2006a,liu:2007a} here we find that re-exponentiation of the short-time quadratic decay shoulder is never justified.  In the perturbative regime $b\gtrsim A$, the system is Markovian,\cite{coish:2008a} being well-described by a single-exponential decay.  As the electron Zeeman splitting is lowered to $b\lesssim A$, we find strong non-Markovian effects (sustained oscillations and multiple decay rates), which once again invalidate re-exponentiation of the short-time decay shoulder.

In addition to recovering previous results, we have found qualitatively new behavior, including modulations of the decay envelope and sub-leading corrections to the decoherence rate for the dominant exponential decay.  Our calculation gives an interesting non-monotic dependence of the decoherence rate $1/T_2$ on magnetic field.  These two results (envelope modulations and a non-monotonic dependence of the decoherence rate on magnetic field, both of which should be readily accessible in experiment) are not recovered in dynamics under a leading-order effective Hamiltonian, suggesting caution should be exercised in interpreting results of high-order expansions involving the effective Hamiltonian.

\section{Acknowledgments}
We acknowledge funding from QuantumWorks, an Ontario PDF, the CIFAR JFA, NSERC, the Swiss NSF, NCCR Nanoscience Basel, and JST ICORP.  WAC and DL gratefully acknowledge the hospitality of the Kavli Institute for Theoretical Physics, where much of this work was completed.

\appendix

\section{Self-energy expansion}\label{sec:Self-energy-expansion}

Here we give the explicit self-energy up to fourth order in $V_{\mathrm{ff}}$.
The full self-energy superoperator is given by $\mathsf{\Sigma}(s)=\mathsf{\Sigma}^{(2)}(s)+\mathsf{\Sigma}^{(4)}(s)+O(V_{\mathrm{ff}}^{6})$,
where \begin{eqnarray}
\Sigma^{(2)}(s) & = & -i\mathrm{Tr}\left(S_{+}\mathsf{L}_{V}\mathsf{G}(s)\mathsf{L}_{V}S_{-}\rho_I(0)\right),\\
\mathsf{G}(s) & = & \frac{1}{s+i\mathsf{L}_{0}},\\
\mathsf{L}_{0}O & = & \commute{H_{0}}{O},\\
\mathsf{L}_{V}O & = & \commute{V_{\mathrm{ff}}}{O},\end{eqnarray}
 and the fourth-order result is

\begin{multline}
\Sigma^{(4)}(s)=i\mathrm{Tr}\left[S_{+}\left(1-iL_{0}Q\mathsf{G}(s)\right)L_{V}\mathsf{G}(s)L_{V}Q\right.\\
\left.\times\mathsf{G}(s)L_{V}\mathsf{G}(s)L_{V}S_{-}\rho_I(0)\right].
\end{multline}
More explicitly, using Eq. (\ref{eq:RhoIZero}) for the initial nuclear state we find $\Sigma^{(p)}(s)=\sum_l\rho_{ll}\Sigma^{(p)}_l(s)$, where
\begin{multline}
\Sigma^{(2)}_l(s)=-\frac{i}{4}\sum_{k}\left(\frac{\left[h^{-}\right]_{n_lk}\left[h^{+}\right]_{kn_l}}{s+i\delta\omega_{kn}-i\omega_{kn_l}^{I}}\right.\\
+\left.\frac{\left[h^{+}\right]_{n_lk}\left[h^{-}\right]_{kn_l}}{s+i\delta\omega_{kn}+i\omega_{kn_l}^{I}}\right),\label{eq:Sigpp2Explicit}
\end{multline}

\begin{widetext}

\begin{multline}
\Sigma^{(4)}_l(s)=\frac{i}{16}\sum_{k_{1}k_{2}k_{3}}\left\{ \left[h_{+}\right]_{n_lk_{1}}\left[h_{-}\right]_{k_{1}k_{2}}\left[h_{+}\right]_{k_{2}k_{3}}\left[h_{-}\right]_{k_{3}n_l}\left[G_{\uparrow}\right]_{k_{1}n_l}\left[G_{+}\right]_{k_{2}n_l}\left[G_{\uparrow}\right]_{k_{3}n_l}\left(1-\delta_{n_lk_{2}}\right)\right.\\
+\left[h_{-}\right]_{n_lk_{1}}\left[h_{+}\right]_{k_{1}k_{2}}\left[h_{-}\right]_{k_{2}k_{3}}\left[h_{+}\right]_{k_{3}n_l}\left[G_{\downarrow}\right]_{n_lk_{3}}\left[G_{+}\right]_{n_lk_{2}}\left[G_{\downarrow}\right]_{n_lk_{1}}\left(1-\delta_{n_lk_{2}}\right)\\
+\left[h_{-}\right]_{n_lk_{1}}\left[h_{+}\right]_{k_{1}k_{2}}\left[h_{+}\right]_{k_{2}k_{3}}\left[h_{-}\right]_{k_{3}n_l}\left(1-i\left(\left[L_{0}^{+}\right]_{k_{2}k_{2}}-\left[L_{0}^{+}\right]_{n_ln_l}\right)\left[G_{+}\right]_{k_{2}k_{2}}\right)\\
\times\left[\left({{\left[G_{\downarrow}\right]_{k_{2}k_{1}}\left[G_{+}\right]}_{k_{2}n_l}\left[G_{\uparrow}\right]}_{k_{3}n_l}+{{\left[G_{\uparrow}\right]_{k_{3}k_{2}}\left[G_{+}\right]}_{{n_lk}_{2}}\left[G_{\downarrow}\right]}_{n_lk_{1}}\right)\left(1-\delta_{n_lk_{2}}\right)\right.\\
\left.\left.+\left(\left[G_{\uparrow}\right]_{k_{3}k_{2}}+\left[G_{\downarrow}\right]_{k_{2}k_{1}}\right)\left[G_{-}\right]_{k_{3}k_{1}}\left(\left[G_{\uparrow}\right]_{k_{3}n_l}+\left[G_{\downarrow}\right]_{n_lk_{1}}\right)\right]\right\} .\label{eq:Sigpp4Explicit}
\end{multline}
\end{widetext}

Here, we denote matrix elements $\left[h^{\pm}\right]_{nk}=\bra{n}h^{\pm}\ket{k}$.
Further, $\delta\omega_{nk}=\frac{1}{2}\left(h_{n}^{z}-h_{k}^{z}\right)$
and $\omega_{nk}^{I}=\bra{n}\omega^{I}\ket{n}-\bra{k}\omega^{I}\ket{k}$,
where $\omega^{I}=b\sum_{k}\gamma_{k}I_{k}^{z}$, and we have introduced
\begin{equation}
\left[G_{\alpha}\right]_{kk^{\prime}}=\frac{1}{s+i\left[L_{0}^{\alpha}\right]_{kk^{\prime}}},
\end{equation}

with\begin{eqnarray}
L_{0}\ket{\downarrow}\bra{\uparrow}\ket{k}\bra{k^{\prime}} & = & \left[L_{0}^{+}\right]_{kk^{\prime}}\ket{\downarrow}\bra{\uparrow}\ket{k}\bra{k^{\prime}},\\
L_{0}\ket{\uparrow}\bra{\downarrow}\ket{k}\bra{k^{\prime}} & = & \left[L_{0}^{-}\right]_{kk^{\prime}}\ket{\uparrow}\bra{\downarrow}\ket{k}\bra{k^{\prime}},\\
L_{0}\ket{\uparrow}\bra{\uparrow}\ket{k}\bra{k^{\prime}} & = & \left[L_{0}^{\uparrow}\right]_{kk^{\prime}}\ket{\uparrow}\bra{\uparrow}\ket{k}\bra{k^{\prime}},\\
L_{0}\ket{\downarrow}\bra{\downarrow}\ket{k}\bra{k^{\prime}} & = & \left[L_{0}^{\downarrow}\right]_{kk^{\prime}}\ket{\downarrow}\bra{\downarrow}\ket{k}\bra{k^{\prime}}.
\end{eqnarray}
The dominant contributions to the self-energy occur at high frequency ($s\simeq i\omega_n$) in the lab frame. For $\left|s-i\omega_{n}\right|\ll\omega_{n}$, and $\omega_{n}\gg\delta\omega_{nk}$, $\omega_{n}\gg\omega_{n_lk}^{I}$, we have:
\begin{equation}
\left[G_{\uparrow}\right]_{kn_l}\approx\left[G_{\downarrow}\right]_{kn_l}=\frac{1}{i\omega_{n}}\left(1+O\left[\frac{A}{N\omega_n}\right]\right),
\end{equation}
which allows us to approximate Eqs. (\ref{eq:Sigpp2Explicit}),(\ref{eq:Sigpp4Explicit})
by their high-frequency forms for a uniformly polarized system.  We additionally go to the rotating frame; from the definition of $\tilde{\Sigma}$ in Eq. \eqref{eq:SigmaRotatingFrame}, we have 
\begin{equation}
\Sigma(s+i \omega_n) = \tilde{\Sigma}(s-i \Delta \omega),
\end{equation}
which gives
\begin{equation}
\tilde{\Sigma}^{(2)}(s-i\Delta \omega)\approx-\frac{1}{4\omega_{n}}\sum_{i}\nu_{i}\left(c_{+}^{i}+c_{-}^{i}\right)\sum_{k}\left(A_{k}^{i}\right)^{2}.\label{eq:Sigpp2HighFreq}
\end{equation}
In the above expression, the sum over $i$ indicates a sum over different
nuclear-spin isotopes with abundances $\nu_{i}$ and hyperfine coupling
constants $A_{k}^{i}$.  The high-frequency form of the fourth-order self-energy is then:
\begin{widetext}
\begin{multline}
\tilde{\Sigma}^{(4)}(s-i \Delta\omega)=\frac{-i}{16\omega_{n}^{2}}\sum_{ij}\nu_{i}\nu_{j}c_{-}^{i}c_{+}^{j}\sum_{k_{1}k_{2}}\left(A_{k_{1}}^{i}\right)^{2}\left(A_{k_{2}}^{j}\right)^{2}\left[\frac{1}{s+ix_{12}^{ij}-i\gamma_{ij}}+\frac{1}{s-ix_{12}^{ij}-i\gamma_{ij}}\right.\\
\left.+\left(\frac{s}{s+i2\gamma_{ij}}\right)\left(\frac{2}{s+i2x_{12}^{ij}}-\frac{1}{s+ix_{12}^{ij}+i\gamma_{ij}}\right)+\left(\frac{s}{s-i2\gamma_{ij}}\right)\left(\frac{2}{s+i2x_{12}^{ij}}-\frac{1}{s+ix_{12}^{ij}-i\gamma_{ij}}\right)\right],\label{eq:Sigpp4HighFreq}
\end{multline}
\end{widetext}
where $x_{12}^{ij}=(A_{k_{1}}^{i}-A_{k_{2}}^{j})/2$,
$\gamma_{ij}=b(\gamma_{i}-\gamma_{j})$, and the coefficients $c_{\pm}^{i}$
are:
\begin{equation}
c_{\pm}^{i}=I_{i}(I_{i}+1)-\left\langle \left\langle m(m\pm1)\right\rangle \right\rangle,
\end{equation}
With the average $\langle\langle\cdots\rangle\rangle$ defined in Eq. (\ref{eq:avg-definition}). For a homonuclear system, we have $\gamma_{ij}=0$ and $x_{12}^{ij}=x_{12}=(A_{k_1}-A_{k_2})/2$,
and replace $\sum_{ij} \nu_i \nu_j \rightarrow 1$.  In this case, assuming a uniformly-polarized nuclear-spin system, the self-energy is given simply by
\begin{equation}\label{eq:self-energy-discrete}
 \tilde{\Sigma}^{(4)}(s-i \Delta\omega) = -i\frac{c_+c_-}{4\omega_n^2}\sum_{k,k'}\frac{A_k^2 A_{k'}^2}{s-i(A_k-A_{k'})}.
\end{equation}
This self-energy differs from that found previously at leading order in an effective-Hamiltonian treatment,\cite{coish:2008a} where the Lamb shift $\Delta\omega$ is incorporated directly into the bare precession frequency $\omega_n$.  In addition, we stress that the more general self-energy for a heteronuclear system (Eq. \eqref{eq:Sigpp4HighFreq}) is not recovered with the effective Hamiltonian (compare with Eq. (C19) of Ref. \onlinecite{coish:2008a}). 

Assuming a two-dimensional parabolic quantum dot (with Gaussian envelope function) leads to coupling constants $A_k = (A/N)e^{k/N}$ (see, e.g., Ref. \onlinecite{coish:2008a}).  Performing the continuum limit, i.e., replacing $\sum_{k_1, k_2} \rightarrow \int d k_1 dk_2$, and evaluating the resulting energy integrals, we arrive at Eq. \eqref{eq:sigma4} of the main text.

\section{Higher-order corrections}\label{sec:higher-order}
All results in this article are valid up to fourth order in the electron-nuclear flip-flop terms $V_\mathrm{ff}$.  As the electron Zeeman splitting is lowered from $b\gg A$, higher-order corrections to the self-energy may become relevant.  In this Appendix, we give explicit conditions under which higher-order corrections may be neglected, even for $b\sim A$.  As in Appendix \ref{sec:Self-energy-expansion}, the self-energy at any order may be approximated by its high-frequency form (at $s\simeq i\omega_n$) whenever $A/Nb\ll 1$.  This allows for a significant simplificaiton in the high-order expansion in terms of $V_\mathrm{ff}$.  With corrections to the self-energy that are smaller by factors of order $1/N\ll 1$ and $A/Nb\ll 1$, we find the high-frequency form of the self-energy to be given by
\begin{equation}\label{eq:self-energy-simplified}
 \Sigma(s) \simeq-i\mathrm{Tr}_I\left[\left(\mathsf{G}_+^{-1}\mathsf{Q}^{-1}+\frac{i}{2}\mathsf{L}^+_\omega\right)\mathsf{\sigma}\frac{1}{1+\mathsf{\sigma}}\rho_I(0)\right],
\end{equation}
where we have introduced 
\begin{eqnarray}
 \mathsf{G}_+ &=& \frac{1}{s-\frac{i}{2}\mathsf{L}^+_\omega},\\
\mathsf{\sigma} &=&  -\frac{i\mathsf{Q}}{4\omega_n}\mathsf{G}_+\left(\mathsf{H}_L+\mathsf{H}_R\right),
\end{eqnarray}
defined in terms of the superoperators (which act on an arbitrary operator $\mathcal{O}$):
\begin{eqnarray}
 \mathsf{H}_L\mathcal{O} &=& h_+h_-\mathcal{O},\\
 \mathsf{H}_R\mathcal{O} &=& \mathcal{O}h_-h_+,\\
 \mathsf{L}^+_\omega\mathcal{O} &=& \left\{\omega,\mathcal{O}\right\},
\end{eqnarray}
where $\left\{,\right\}$ indicates an anticommutator.

The high-frequency form of the self-energy can now be found directly from Eq. (\ref{eq:self-energy-simplified}) with a more moderate constraint on the electron Zeeman splitting ($A/Nb\ll 1$).  A direct evaluation of Eq. (\ref{eq:self-energy-simplified}) at arbitrary order and resummation is non-trivial, but we can generate arbitrary higher-order terms with the geometric series:
\begin{equation}
\frac{1}{1+\sigma}=1-\sigma+\sigma^2-\cdots.
\end{equation}

Every factor of $\sigma$ is associated with a nuclear-spin pair flip, giving rise to a factor of $c_+$ or $c_-$, which depend on the nuclear polarization, and a factor of the small parameter $A/\omega_n$.  The term at $(2k)^\mathrm{th}$ order in $V_\mathrm{ff}$ contains $k$ factors of $\sigma$, and consequently $k$ powers of $A/\omega_n$.  This suggests that the sixth-order self-energy can in general give corrections of order $\sim \left(A/\omega_n\right)^3$, which may modify the sub-leading corrections of this size given by the Markovian decay formula (Eq. (\ref{eq:homonucT2})).  However, by direct calculation using the above expansion, we find the leading contributions to the Markovian decay rate at sixth order to be
\begin{equation}
 -\mathrm{Im}\Sigma^{(6)}(s=i\omega_n+0^+) = \mathcal{O}\left[\left(\frac{A}{\omega_n}\right)^4\right].
\end{equation}
Furthermore, we find that the $\Sigma^{(6)}$ corrections do not lead to a broadening of the continuum band. The first non-vanishing corrections to the Markovian decay rate which do lead to a broadening of the continuum band contain two nuclear-spin pair-flip excitations.  These terms occur first at eighth order in $V_\mathrm{ff}$ and are suppressed by the factor $\left(c_+c_-\right)^2$--which is smaller than the fourth-order corrections by a factor $c_+c_-$ for a polarized nuclear-spin system (e.g. $c_+c_-\propto (1-p^2)$ for nuclear spin $I=1/2$).  This result demonstrates that the qualitative decrease in the decoherence rate at low electron Zeeman splitting shown in Figs. \ref{fig:gaasrates}, \ref{fig:ingaas-rates}, and \ref{fig:ingaas-doping} will not be significantly modified by higher-order corrections, at least in the case of a large polarization, where perturbation theory still applies at a smaller value of the electron Zeeman splitting.

\bibliographystyle{apsrev}
\bibliography{fid}

\end{document}